\documentclass[letterpaper, 10 pt, conference]{ieeeconf}  % Comment this line out
\IEEEoverridecommandlockouts                              % This command is only
\usepackage{epsfig} 
\usepackage{subfig}
\usepackage{graphicx}    % include this line if your document contains figures
\usepackage{epstopdf} 
\usepackage{amsmath} % assumes amsmath package installed
\usepackage{amssymb}  % assumes amsmath package installed
\usepackage{multirow}
\usepackage[dvipsnames]{xcolor}
\usepackage{soul}
\usepackage[cmintegrals]{newtxmath}
\usepackage{float}
\usepackage{mathrsfs}
\usepackage{algorithmic}
\usepackage{algorithm}
\usepackage{subdepth}

\renewcommand{\ge}{\geq}
\newcommand\e{\mathrm{e}}
\newcommand\E{\mathbb{E}}

\renewcommand{\le}{\leq}
\newcommand{\R}{\mathbb{R}}
\newcommand\tr{\mathop{\mathrm{tr}}}

\newcommand\T{\rm T}
\newcommand{\mR}
{\mathbb{R}}
\usepackage{multirow}
\usepackage{color}

\DeclareMathAlphabet\mathbfcal{OMS}{cmsy}{b}{n}

\definecolor{cover}{RGB}{72,0,255}
\definecolor{gold}{RGB}{204,163,0}
\definecolor{darkblue}{RGB}{0,21,120}
\definecolor{rred}{RGB}{225,81,18}
\definecolor{dred}{RGB}{170,60,10}
\definecolor{bblue}{RGB}{0,115,189}
\definecolor{ggreen}{RGB}{0,160,100}
\definecolor{dgreen}{RGB}{0,99,61}
\definecolor{lblue}{RGB}{77,191,237}

\definecolor{d2red}{RGB}{215,48,39}
\definecolor{d2orange}{RGB}{252,141,89}
\definecolor{d2yellow}{RGB}{254,224,144}
\definecolor{d2lblue}{RGB}{224,243,248}
\definecolor{d2blue}{RGB}{145,191,219}
\definecolor{d2dblue}{RGB}{69,117,180}

\newtheorem{Theorem}{Theorem}
\newtheorem{corollary}{Corollary}

\newtheorem{Prop}{Proposition}
\newtheorem{Remark}{Remark}
\newtheorem{Assumption}{Assumption}

\title{\LARGE \bf An $H_2$-norm approach to performance analysis of networked control systems under multiplicative routing transformations
}

\author{{Ruslan Seifullaev$^{1}$ and Andr\'{e} M. H. Teixeira$^{1}$}% <-this % stops a space
\thanks{This work has been submitted to the IEEE for possible publication. Copyright may be transferred without notice, after which this version may no longer be accessible.}
\thanks{\mbox{$^1$}Division of  Systems and Control, Department of Information Technology, Uppsala University, Sweden.}
\thanks{ Email addresses:
        {\tt\footnotesize \{ruslan.seifullaev, andre.teixeira\} @it.uu.se}}
}

\begin{document}

\maketitle
\thispagestyle{empty}
\pagestyle{empty}

%%%%%%%%%%%%%%%%%%%%%%%%%%%%%%%%%%%%%%%%%%%%%%%%%%%%%%%%%%%%%%%%%%%%%%%%%%%%%%%%
\begin{abstract}
This paper investigates the performance of networked control systems subject to multiplicative routing transformations that alter measurement pathways without directly injecting signals. Such transformations, arising from faults or adversarial actions, modify the feedback structure and can degrade performance while remaining stealthy. An $H_2$-norm framework is proposed to quantify the impact of these transformations by evaluating the ratio between the steady-state energies of performance and residual outputs. Equivalent linear matrix inequality (LMI) formulations are derived for computational assessment, and analytical upper bounds are established to estimate the worst-case degradation. The results provide structural insight into how routing manipulations influence closed-loop behavior and reveal conditions for stealthy multiplicative attacks.
\end{abstract}

%%%%%%%%%%%%%%%%%%%%%%%%%%%%%%%%%%%%%%%%%%%%%%%%%%%%%%%%%%%%%%%%%%%%%%%%%%%%%%%%
%\section{Introduction}

\section{Introduction}
Networked control systems (NCSs) integrate spatially distributed sensors, actuators, and controllers through shared communication channels. This architecture enables flexibility and scalability in applications such as industrial automation, power systems, and transportation networks. However, network-induced constraints and imperfections can significantly affect closed-loop stability and performance. Moreover, the use of open or wireless communication channels exposes NCSs to faults and adversarial actions that manipulate transmitted data or disrupt the underlying network structure.

Beyond random network imperfections, NCSs are increasingly exposed to deliberate manipulations of the communication layer. Faults caused by hardware degradation, software errors, or compromised firmware can alter transmitted signals, leading to degraded performance or instability. Of particular concern are attacks that modify signal routing or interconnections among networked components, as these can remain stealthy while reshaping the closed-loop behavior in a structured yet deceptive manner. Unlike additive or replay-type attacks, which inject or replace data samples, multiplicative or routing attacks modify how measurement signals are distributed, scaled, or permuted before reaching the controller. Such transformations can be implemented, for instance, by altering network routing tables, reconfiguring wireless links, or modifying sensor fusion weights at the firmware level. Even without adding external energy, these attacks can distort the system’s feedback pathways, degrade performance, and affect detectability.

The literature on secure control has extensively studied various classes of network attacks \cite{Cardenas08,Sandberg15,22_Teixeira25}, including denial-of-service (DoS) \cite{DePersis,23_Dolk17,Seifullaev2024}, deception and false-data injection attacks \cite{ZHAO2020109128,Liu11}, replay \cite{Mo15}, and zero-dynamics attacks \cite{Pasqualetti13,Smith15}. These works have led to a variety of fault detection, isolation, and resilient control strategies \cite{24_Mousavinejad}. However, while additive and injection-type attacks have been well characterized, the study of multiplicative manipulations
%—those that reconfigure measurement routing without directly injecting spurious energy—
has received significantly less attention. Such attacks are particularly concerning because, unlike additive manipulations that require the injection of external energy, multiplicative transformations can modify feedback pathways with minimal or no energy expenditure, while still degrading closed-loop performance and potentially evading energy-based anomaly detectors.

A common framework for assessing the impact of faults and attacks relies on input-output system norms such as the $H_2$ norm, which provides a natural quantitative measure of how structural changes in communication pathways affect overall system performance and residual energy used for detection  \cite{Shames17}.
% \cite{Zhou96}.
Since the detectability of multiplicative faults inherently depends on the external excitation of the system, the $H_2$-norm, which captures steady-state energy propagation under stochastic or persistent excitation, offers a physically meaningful performance and detectability metric.
In contrast to  $H_\infty$-based metrics, which focus on additive attacks, the present work investigates the structural effects of multiplicative routing transformations.

In this paper, we study linear NCSs subject to multiplicative attacks in which the adversary gains access to the sensor network and replaces the measurement vector $y$ with a transformed version $\tilde y=Ry$, where $R$ is an unknown routing matrix. This model captures a broad range of attack scenarios, including selective signal blocking, gain manipulation, and channel permutation. The resulting closed-loop system remains linear but with dynamics that depend explicitly on $R$, which affects both stability and performance. Assuming that the adversary maintains system stability to avoid immediate detection, we analyze how different routing configurations influence the trade-off between performance degradation and residual energy generation.

The main contributions of this paper are summarized as follows:
\begin{itemize}
\item 
We formulate the attack impact as an optimization problem that maximizes the ratio between the steady-state energies of the performance and residual outputs for a system excited by white-noise, and we show that this ratio admits an exact representation in terms of the system’s 
$H_2$-norms.

\item
For a given routing transformation, we derive equivalent LMI conditions that allow efficient computation of the $H_2$-based performance ratio. Furthermore, we establish analytical upper bounds on the ratio under small and general perturbations of the routing matrix $R$, providing tractable estimates of the worst-case performance degradation.

\item
We reveal how specific structural alignments between the attack-induced routing transformation and the system’s closed-loop modes can amplify the performance output while keeping residuals small, characterizing conditions for stealthy multiplicative attacks.
\end{itemize}

\textbf{Notations:} Throughout the paper, 
$\Vert G\Vert_{H_2}=\sqrt{\frac{1}{2\pi}\int_{-\infty}^\infty \tr\{G(i\omega)G^*(i\omega)\}d\omega}$ 
denotes the $H_2$ norm of a linear system with the transfer matrix $G(s)$. 
For a real matrix $M$, 
$\sigma_{\rm min}(M)$ and $\sigma_{\rm max}(M)$ denote its minimum and maximum singular values, respectively, 
while $\lambda_{\rm min}(M)$ and $\lambda_{\rm max}(M)$ denote the minimum and maximum eigenvalues when $M$ is square. 
The condition number of a nonsingular matrix $M$ is defined as 
$\kappa(M)=\frac{\sigma_{\rm max}(M)}{\sigma_{\rm min}(M)}$. 
For a real-valued matrix $M$, the spectral 2-norm and Frobenius norm are given by 
$\Vert M\Vert_2=\sigma_{\rm max}(M)$ and 
$\Vert M\Vert_F=\sqrt{\tr\{M^{\T}M\}}$, respectively. 
Note that if a  symmetric matrix $P$ is positive definite, then 
$\kappa(P)=\frac{\lambda_{\rm max}(P)}{\lambda_{\rm min}(P)}=\frac{\Vert P\Vert_2}{\lambda_{\rm min}(P)}$.

The rest of the paper is organized as follows. Section~II formulates the problem setup and introduces the closed-loop model under multiplicative attacks. Section~III develops the $H_2$-norm representation of the performance ratio and its equivalent LMI formulations. Section~IV presents analytical upper bounds and worst-case estimates for the attack impact. A numerical example demonstrating the influence of routing transformations on closed-loop performance is provided in Section~V. Finally, Section~VI concludes the paper.

\section{Problem statement}
We consider the following linear system:
\begin{equation}\label{sys_lin_per}
\begin{aligned}
\dot x_{\rm p}(t) &= A_{\rm p}x_{\rm p}(t)+B_{\rm p}u(t)+B_ww_{\rm p}(t),\\ 
y_{\rm m}(t) &=C_{\rm m,o}x_{\rm p}(t),\quad y_{\rm p}(t) =C_{\rm p,o}x_{\rm p}(t)+D_{\rm p,o}u(t),
\end{aligned}
\end{equation}
where $x_{\rm p}(t) \in \mR^{n_x}$ is the state vector, $u(t) \in \mR^{n_u}$ is the control input, $y_{\rm m}(t) \in \mR^{n_y}$ is the measurement output, $y_{\rm p} \in \mR^{n_{y_{\rm p}}}$ is the performance output, and $A_{\rm p}$, $B_{\rm p}$, $B_w$, $C_{\rm m,o}$, $C_{\rm p,o}$ and $D_{\rm p,o}$ are the matrices of appropriate dimensions, the pair $(A_{\rm p}, B_{\rm p})$ is controllable and the pair $(A_{\rm p}, C_{\rm m,o})$ is observable. The input $w_{\rm p}(t)\in\mR^{n_{x}}$ is a white noise with unit intensity, i.e., $S_w(\omega)\equiv I$, where $S_w(\omega)$ is the spectral density of $w$.

\begin{figure}
\begin{center}
\includegraphics[scale=1.5]{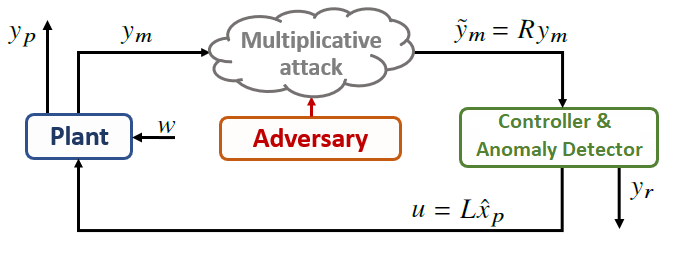}
\end{center}
\caption{Multiplicative attack}\label{Mult}
\end{figure}

Assume that the system is under a multiplicative attack, where an adversary can access the sensor measurements and replace the vector $y_{\rm m}(t)$ with 
$$
\tilde y_{\rm m}(t)=Ry_{\rm m}(t),
$$
where $R\in \R^{n_y \times n_y}$ is some matrix, see Fig.~\ref{Mult}.
Therefore, instead of the true measurement output $ y_{\rm m}$, the controller is provided with the modified signal $\tilde y_{\rm m}$. 

%\subsection{Observer-based control}
To estimate the state vector $x_{\rm p}(t)$ and design a feedback based on it, we use an observer-based state feedback controller. Additionally, we use its residual output $y_{\rm r}(t)$, the difference between the measured and predicted outputs,  for an anomaly detector, which raises an alarm if the residual becomes abnormally high.

We consider the following controller structure: 
\begin{equation}\label{obs_lin}
\begin{aligned}
\dot{\hat x}_{\rm p}(t) &= A_{\rm p}\hat x_{\rm p}(t)+B_{\rm p}u(t)+B_{\hat w}\hat w_{\rm p}(t)+Ky_{\rm r}(t),\\
u(t) &= L\hat x_{\rm p}(t),\,\, \hat y_{\rm m}(t) =C_{\rm m,o}\hat x_{\rm p}(t),\,\,
y_{\rm r}(t) = \tilde y_{\rm m}(t)-\hat y_{\rm m}(t),
\end{aligned}
\end{equation}
where $B_{\hat w}\in\mR^{n_x\times n_x}$, $\hat w_{\rm p}(t)$ is white noise with unit intensity, $K\in\mR^{n_x\times n_y}$ and $L\in\mR^{n_u\times n_x}$ are the observer and controller gains, respectively.
The closed-loop system \eqref{sys_lin_per}, \eqref{obs_lin} then becomes
\begin{equation}\label{sys_lin_gen_cl}
\begin{aligned}
\dot x(t) &= A_Rx(t)+Bw(t),\\ 
y_{\rm r}(t) &=C_{{\rm r}, R}x(t),\quad y_{\rm p}(t) =C_{\rm p}x(t),
\end{aligned}
\end{equation}
where 
$
x=  \begin{bmatrix}
x_{\rm p}\\
e\\
\end{bmatrix}= \begin{bmatrix}
x_{\rm p}\\
x_{\rm p}-\hat x_{\rm p}\\
\end{bmatrix}, w=  \begin{bmatrix}
w_{\rm p}\\
\hat w_{\rm p}\\
\end{bmatrix}
$ and
\begin{equation*}
\begin{aligned}
 A_R & = \begin{bmatrix}
A_{\rm p}+B_{\rm p}L&-B_{\rm p}L\\
K(I-R)C_{\rm m,o}&A_{\rm p}-KC_{\rm m,o}\\
\end{bmatrix},\,\, B= \begin{bmatrix} B_w&0\\B_w&-B_{\hat w}\\\end{bmatrix},\\
C_{{\rm r}, R} & = \left[ 
(R-I)C_{\rm m,o},\,\, C_{\rm m,o}
\right], \,\, C_{\rm p}  = \left[ 
C_{\rm p,o}+D_{\rm p,o}L,\, -D_{\rm p,o}L
\right]\!.
\end{aligned}
\end{equation*}
\begin{Remark}
Note that if $R=I$ (no attack), then the eigenvalues of the closed-loop system \eqref{sys_lin_gen_cl} are those of the matrices $A_{\rm p}+B_{\rm p}L$ (state feedback) and $A_{\rm p}-KC_{\rm m,o}$ (observer). Since \eqref{sys_lin_per} is both controllable and observable, it is always possible to design the gain matrices $L$ and $K$ to ensure the stability of the closed-loop system. However, for arbitrary $R$, stability can be compromised.   
In other words, the adversary can design the matrix $R$ such that $A_R$ becomes unstable. Nonetheless, making the system completely unstable would likely be easily detected\footnote{In this paper, we do not address the specific design of anomaly detectors; however, a typical anomaly detector raises an alarm if the energy of $y_{\rm r}$ exceeds a certain threshold, e.g.,
%\begin{equation}\label{detector}
$\E\left\{y_{\rm r}^{\T}(t)y_{\rm r}(t)\right\}> \varepsilon_{\rm tr}$.
%\end{equation}
For instance, under the assumption that the input noise $w$ follows a Gaussian distribution, the corresponding residual signal energy follows a  $\chi^2$ distribution. In this case, the threshold can be chosen based on its expected value. See, for example,  
\cite{Bai2017} and \cite{Guo2019} for further details.
}. Therefore, it is reasonable to restrict $R$ to the class of the matrices for which the matrix $A_R$ remains stable.
\end{Remark}
\begin{Assumption}
We assume that $R\in  \mathcal{R}_{\rm s}^{n_y}$, where
$$
\mathcal{R}_{\rm s}^{n_y}=\left\{ R\in \R^{n_y\times n_y}: A_R\,\,\mbox{is Hurwitz}\right\}.
$$
\end{Assumption}

The adversary aims to manipulate the sensor network architecture and signal routing so as to maximize the energy of the performance output $y_{\rm p}$ while keeping the energy of the residual output $y_{\rm r}$ small. This trade-off can be formulated as the problem of finding the attack matrix $R$ that maximizes the ratio between the energies $y_{\rm p}$ and $y_{\rm r}$.
\begin{equation}\label{max_prob01}
\sup\limits_{R\in \mathcal{R}_{\rm s}^{n_y}, x(0)=0}\lim_{t\to\infty}\frac{\E\left\{y_{\rm p}^{\T}(t)y_{\rm p}(t)\right\}}{\E\left\{y_{\rm r}^{\T}(t)y_{\rm r}(t)\right\}}.
\end{equation}
\begin{Remark}
This work focuses exclusively on the structural effect of multiplicative (routing) attacks represented by the matrix $R$. Such attacks do not inject additional energy into the system but modify the measurement channels by blocking, scaling, or permuting sensor signals. Hence, they can be interpreted as a preparatory stage that reshapes the signal routing before any active, energy-consuming attack takes place.

To isolate and quantify this structural influence, we consider the system driven only by process noise $w(t)$. The stochastic excitation serves as a persistent, energy-neutral source that allows us to evaluate the steady-state statistical properties of the closed-loop system under different configurations of $R$. The resulting performance–residual trade-off can be naturally expressed as a ratio \eqref{max_prob01}, which measures the steady-state energy gains from $w$ to the outputs.
\end{Remark}

\begin{Remark}
We also add a small independent noise term in the observer dynamics to regularise the estimator. This modification can be interpreted as adding small perturbations in the estimator, which makes the closed-loop system well-posed, 
%excites otherwise unobserved state directions, 
and, from a practical perspective, acts as a protective mechanism by exciting all state directions, thereby enhancing the detectability of stealthy routing manipulations.
\end{Remark}

{\it Problem formulation}: The goal of this paper is to characterize the effect of multiplicative attacks on system performance by deriving and analyzing an exact $H_2$-norm representation of the optimization problem \eqref{max_prob01}. 
This $H_2$-based formulation provides a physically meaningful characterization of how the routing matrix $R$ alone affects performance and detectability.

%Thus, it is reasonable for the adversary to aim at degrading system performance without triggering detection.

\section{$H_2$-norm representation}
To analyze the performance degradation under multiplicative attacks, we reformulate the optimization problem~\eqref{max_prob01} in terms of the system's $H_2$-norms. This representation connects the steady-state signal energies to the underlying system dynamics and allows deriving tractable expressions for the performance ratio.

Define the transfer matrices
$$G_{\rm p}(s)=C_{\rm p}(sI-A_R)^{-1}B, \quad G_{\rm r}(s)=C_{{\rm r}, R}(sI-A_R)^{-1}B.
$$
The following proposition provides an equivalent formulation of the optimization problem~\eqref{max_prob01} as an $H_2$-norm ratio.
\begin{Prop}\label{prop_H2}
Consider the optimization problems
\begin{equation}\label{max_prob2}
\sup\limits_{R\in \mathcal{R}_{\rm s}^{n_y}}\frac{\Vert G_{\rm p}\Vert^2_{H_2}}{\Vert G_{\rm r}\Vert^2_{H_2}},
\end{equation}
and
\begin{equation}\label{max_prob}
\sup\limits_{R\in \mathcal{R}_{\rm s}^{n_y}}\frac{\tr\left\{ C_{\rm p}P_RC_{\rm p}^{\T}\right\}}{\tr\left\{ C_{{\rm r}, R}P_RC_{{\rm r}, R}^{\T}\right\}},
\end{equation}
where $P_R$ is the controllability Gramian over the interval $[0, T]$, i.e.,
\begin{equation}\label{Gram}
P_R = \int_0^T\e^{A_Rt}BB^{\T}\e^{A_R^{\T}t}dt.
\end{equation}
Then the problems \eqref{max_prob01}, \eqref{max_prob2}, and \eqref{max_prob} (for $T\to\infty$) are equivalent.
\end{Prop}
\begin{proof}
See Appendix.\end{proof}

When calculating $H_2$-norms, direct evaluation of the integral in~\eqref{Gram} can be computationally intensive, particularly for large-scale systems. 
Instead of computing the controllability Gramian explicitly, one can evaluate the $H_2$-norms by solving equivalent convex optimization problems with LMI constraints.
If $R$ is fixed, the norms in~\eqref{max_prob} can therefore be obtained as the optimal values of the following LMI-based problem.
\begin{Prop}\label{prop_H2_LMI_Boyd_contr}
For each fixed matrix $R\in \mathcal{R}_{\rm s}^{n_y}$ and $T\to\infty$, the $H_2$-norm $\Vert G_{\rm p}\Vert_{H_2}$
can be computed as $\sqrt{\gamma_{\rm p}}$, where
$\gamma_{\rm p}$ is the optimal value of the following convex optimization problem:
\begin{equation}\label{Opt1_Boyd}
\begin{aligned}
& \inf\limits_{X_{\rm p}, Z_{\rm p}, \gamma_{\rm p}}\gamma_{\rm p}, \quad \mbox{s.t.}\quad \gamma_{\rm p}>0, \quad X_{\rm p}=X_{\rm p}^{\T}>0, \quad \tr\left\{Z_{\rm p}\right\}<\gamma_{\rm p},\\
& \begin{bmatrix}
Z_{\rm p}&C_{\rm p}X_{\rm p}\\
X_{\rm p}C_{\rm p}^{\T}&X_{\rm p}\\
\end{bmatrix}>0, \,\, \begin{bmatrix}
X_{\rm p}A_R^{\T}+A_RX_{\rm p}&B\\
B^{\T}&-I\\
\end{bmatrix}<0.
\end{aligned}
\end{equation}
The expression for $\|G_{\rm r}\|_{H_2}^2$ can be computed analogously.
\end{Prop}
\begin{proof}
See Appendix.
\end{proof}

Proposition~\ref{prop_H2} establishes an exact $H_2$-norm representation of the performance ratio~\eqref{max_prob01} in the form~\eqref{max_prob}, thereby characterizing the worst-case impact of multiplicative attacks on system performance. 
To facilitate further analysis, this optimization problem can be expressed in a more tractable algebraic form, as stated below.

\begin{Prop}\label{prop_H2_2}
For $T\to\infty$, the optimization problem \eqref{max_prob}  is equivalent to
\begin{equation}\label{max_prob3}
\begin{aligned}
&\sup\limits_{R\in \mathcal{R}_{\rm s}^{n_y}, \alpha\ge0}\alpha,\\
&\mbox{s.t.} \,\,\tr\left\{X\left(C_{\rm p}^{\T}C_{\rm p}-\alpha C_{{\rm r}, R}^{\T}C_{{\rm r}, R}\right)\right\}\ge0,\\
&\quad\,\,\, XA_R^{\T}+A_R X+BB^{\T}=0,  \,\,X>0.
\end{aligned}
\end{equation}
\end{Prop}
\begin{proof}
See Appendix.
\end{proof}
\begin{corollary}
The equality constraint in \eqref{max_prob3} can be relaxed to the associated inequality
$$
XA_R^{\T}+A_R X+BB^{\T}<0,  \,\,X>0.
$$
\end{corollary}

\begin{Remark}

The optimization problem~\eqref{max_prob01} (and its equivalent formulations in Propositions~\ref{prop_H2} and~\ref{prop_H2_2}) characterizes the attack directions, parameterized by~$R$, that increase the energy of the performance output while simultaneously reducing the energy of the residual output used for attack detection. In other words, it identifies the worst-case multiplicative attacks that most severely degrade control performance while remaining difficult to detect.
It is also possible to consider the following constrained optimization problem, which explicitly characterizes \emph{stealthy} attacks:
\begin{equation}\label{constr_prob}
\sup\limits_{R\in \mathcal{R}_{\rm s}^{n_y}}\tr\left\{ C_{\rm p}P_RC_{\rm p}^{\T}\right\} \quad \mbox{w.r.t.}\quad \tr\left\{ C_{{\rm r}, R}P_RC_{{\rm r}, R}^{\T}\right\}\le \varepsilon_{\rm tr}.
\end{equation}
\end{Remark}

\section{Worst-Case Estimation}
The $H_2$-norm representation introduced in the previous section provides an exact framework for quantifying the effect of multiplicative attacks through the performance ratio~\eqref{max_prob}. However, the optimization problem remains nonconvex with respect to the attack matrix~$R$, and the exact maximization of the ratio is, in general, analytically intractable. To address this challenge, this section develops upper bounds and tractable estimates of the worst-case performance ratio. 

Let \(A\) denote the nominal closed-loop matrix (corresponding to \(R=I\))
$$
A  = \begin{bmatrix}
A_{\rm p}+B_{\rm p}L&-B_{\rm p}L\\
0&A_{\rm p}-KC_{\rm m,o}\\
\end{bmatrix},
$$
 and let \(P\) be the corresponding controllability Gramian:
\[
A P + P A^{\top} + BB^{\top}=0,\qquad P=P^{\top}>0.
\]
Then, for an attack \(R\), we can write \(A_R=A+\Delta A_R\) and let \(P_R=P+\Delta P_R\).
Here, 
$
\Delta A_R  = \begin{bmatrix}
0&0\\
KE_R&0\\
\end{bmatrix}, \quad E_R=(I-R)C_{\rm m,o}.
$
Similarly, $C_{{\rm r}, R}=C_{{\rm r}}+\Delta C_{{\rm r}, R}$, where $C_{{\rm r}}  = \left[ 
0,\,\, C_{\rm m,o}
\right]$ and $\Delta C_{{\rm r}, R}  = \left[ 
E_R,\,\,0
\right]$.
%Denote also the Frobenius norms \(\|C_{\rm p}\|_F^2=\tr\{C_{\rm p}^{\top}C_{\rm p}\}\), \(\|C_{{\rm r},R}\|_F^2=\tr\{C_{{\rm r},R}^{\top}C_{{\rm r},R}\}\), and the condition number 
%$$
%\kappa(P)=\frac{\lambda_{\max}(P)}{\lambda_{\min}(P)}=\frac{\Vert P\Vert_2}{\lambda_{\min}(P)}.
%$$
%The above equalities hold since the matrix $P$ is symmetric and positive definite.

We will start with the case where the matrix $R$ is close to $I$, such that the Gramian perturbation  $\Delta P_R$ is small. The following theorem theorem establishes an upper bound for the ratio in \eqref{max_prob}.
\begin{Theorem}
\label{thm:ratio_bound}
 Assume that
\begin{equation}\label{cond01}
\|\Delta P_R\|_2 \le \delta_R \,\|P\|_2\qquad\text{for some } \delta_R\in[0,1),
\end{equation}
and that \(\delta_R\) satisfies
\begin{equation}\label{cond}
\delta_R\,\kappa(P) < 1,
\end{equation}
where $\kappa(P)$ is the condition number of $P$.
Then the performance ratio \eqref{max_prob} satisfies
\begin{equation}\label{eq:ratio_bound}
\frac{\tr\{C_{\rm p}P_R C_{\rm p}^{\top}\}}{\tr\{C_{{\rm r},R} P_R C_{{\rm r},R}^{\top}\}}
\;\le\;
\frac{\kappa(P)\left(1+\delta_R\right)}{\,1-\delta_R\kappa(P)\,}\;
\frac{\|C_{\rm p}\|_F^2}{\|E_R\|_F^2+\|C_{{\rm r}}\|_F^2}\;.
\end{equation}
\end{Theorem}
\begin{proof}
For any matrix \(M\) and symmetric \(P_R>0\), the
spectral bounds
$$
\lambda_{\min}(P_R)\,\|M\|_F^2 \le \tr\{ M P_R M^{\rm T}\} \le \|P_R\|_2\,\|M\|_F^2.
$$
hold. Applying these to \(C_{\rm p}\) and \(C_{{\rm r}, R}\) yields
\begin{equation}\label{eq:num_bounds}
\tr\{C_{\rm p}P_R C_{\rm p}^{\rm T}\} \le \|P_R\|_2\,\|C_{\rm p}\|_F^2,
\end{equation}
and
\begin{equation}\label{eq:den_bounds}
\tr\{C_{{\rm r},R}P_R C_{{\rm r},R}^{\rm T}\} \ge \lambda_{\min}(P_R)\,\|C_{{\rm r},R}\|_F^2.
\end{equation}
Note that due to the structure of the matrix $C_{{\rm r},R}$, we have that $\|C_{{\rm r},R}\|_F^2={\|E_R\|_F^2+\|C_{{\rm r}}\|_F^2}$.
Then, using \eqref{eq:num_bounds} and \eqref{eq:den_bounds}, we get the inequality
\begin{equation}\label{eq:rat_bounds}
\frac{\tr\{C_{\rm p}P_R C_{\rm p}^{\rm T}\}}{\tr\{C_{{\rm r},R}P_R C_{{\rm r},R}^{\rm T}\}}
\le
\kappa(P_R)\,\frac{\|C_{\rm p}\|_F^2}{\|E_R\|_F^2+\|C_{{\rm r}}\|_F^2}.
\end{equation}
Therefore, we have to estimate the condition number of $P_R$. Using the  sub-additive property of the norm and \eqref{cond01}, we have
\begin{equation}\label{eq:upp_cond}
\|P_R\|_2 \le \|P\|_2 + \|\Delta P_R\|_2\le (1+\delta_R) \,\|P\|_2.
\end{equation}
Weyl’s inequality for symmetric matrices (see Corollary 4.3.15 in \cite{HornJohnson2012}) gives 
$
\lambda_{\min}(P_R) \ge \lambda_{\min}(P) - \|\Delta P_R\|_2,
$
so that
\begin{equation}\label{eq:low_cond}
\lambda_{\min}(P_R) \ge \lambda_{\min}(P)  - \delta_R\,\|P\|_2=  \lambda_{\min}(P)\left(1-\delta_R\kappa(P)\right).
\end{equation}
Therefore, combining the inequalities \eqref{eq:upp_cond} and \eqref{eq:low_cond}, we get
\[
\kappa(P_R) = \frac{\|P_R\|_2}{\lambda_{\min}(P_R)}
\le  \kappa(P)\,\frac{1+\delta_R}{1-\delta_R\kappa(P)}.
\]
Note that the denominator \(1-\delta\kappa(P)\) is positive due to the assumption \eqref{cond}, so the bound is finite.
Finally, by substituting the upper bound for \(\kappa(P_R)\) into the ratio bound \eqref{eq:rat_bounds}, we immediately obtain \eqref{eq:ratio_bound}.
\end{proof}

\begin{Remark}[On obtaining \(\delta_R\) from \(\Delta A_R\)]\label{Rem2}
The theorem assumes a relative Gramian perturbation bound \(\|\Delta P_R\|_2\le\delta_R\|P\|_2\). Such a \(\delta_R\) can be obtained from bounds on \(\|\Delta A_R\|_2\) when \(E_R\) is sufficiently small, e.g. via the perturbation estimate in \cite{NumEst} (Theorem 8.3.3)
\[
\|\Delta P_R\|_2 \le 2\|H\|_2\,\|\Delta A_R\|_2 \,\Vert P_R\Vert_2,
\quad\text{where }AH+HA^{\rm T}=-I.
\]
Then from \eqref{eq:upp_cond}, we have $$\|\Delta P_R\|_2 \le 2\|H\|_2\,\|\Delta A_R\|_2 \,\left(\|P\|_2 + \|\Delta P_R\|_2\right).$$
Therefore, $\|\Delta P_R\|_2 \le \frac{2\|H\|_2\,\|\Delta A_R\|_2}{1-2\|H\|_2\,\|\Delta A_R\|_2}\|P\|_2.$
Thus, one may take \(\delta_R =\frac{2\|H\|_2\,\|\Delta A_R\|_2}{1-2\|H\|_2\,\|\Delta A_R\|_2}\), and to satisfy \eqref{cond01} and \eqref{cond}, we must require
$
\|\Delta A_R\|_2\le \frac{1}{2\|H\|_2(1+\kappa(P))}.
$
\end{Remark}
%\begin{Remark}[Alternative bound]
%Instead of \eqref{eq:num_bounds}, we can use the following estimate:
%$$
%\tr\{C_p P_R C_p^{\rm T}\} \le \|C_p\|_2^2\,\tr\left\{P_R\right\}\le n_x \|C_p\|_2^2\|P_R\|_2^2. 
%$$
%\end{Remark}

The result of Remark~\ref{Rem2} implies that the upper bound \eqref{eq:ratio_bound} is valid only when \(E_R\) is sufficiently small, i.e., for attacks corresponding to \(R\) close to the identity. However, arbitrary attack matrices may cause large deviations that fall outside this assumption. Therefore, we next consider a more general case without requiring conditions \eqref{cond01} and \eqref{cond}. At the same time, we impose a stability margin on \(A_R\) and exclude attacks that render \(A_R\) slow or marginally stable, as such cases are typically easy to detect. This assumption ensures that the performance ratio remains finite and that the optimization problem \eqref{max_prob} is well-posed. Specifically, we restrict \(R\) such that \(A_R\) is exponentially stable with a guaranteed convergence rate, as formalized in the following theorem.
\begin{figure*}[h!]
\centering
\includegraphics[width=6.5in]{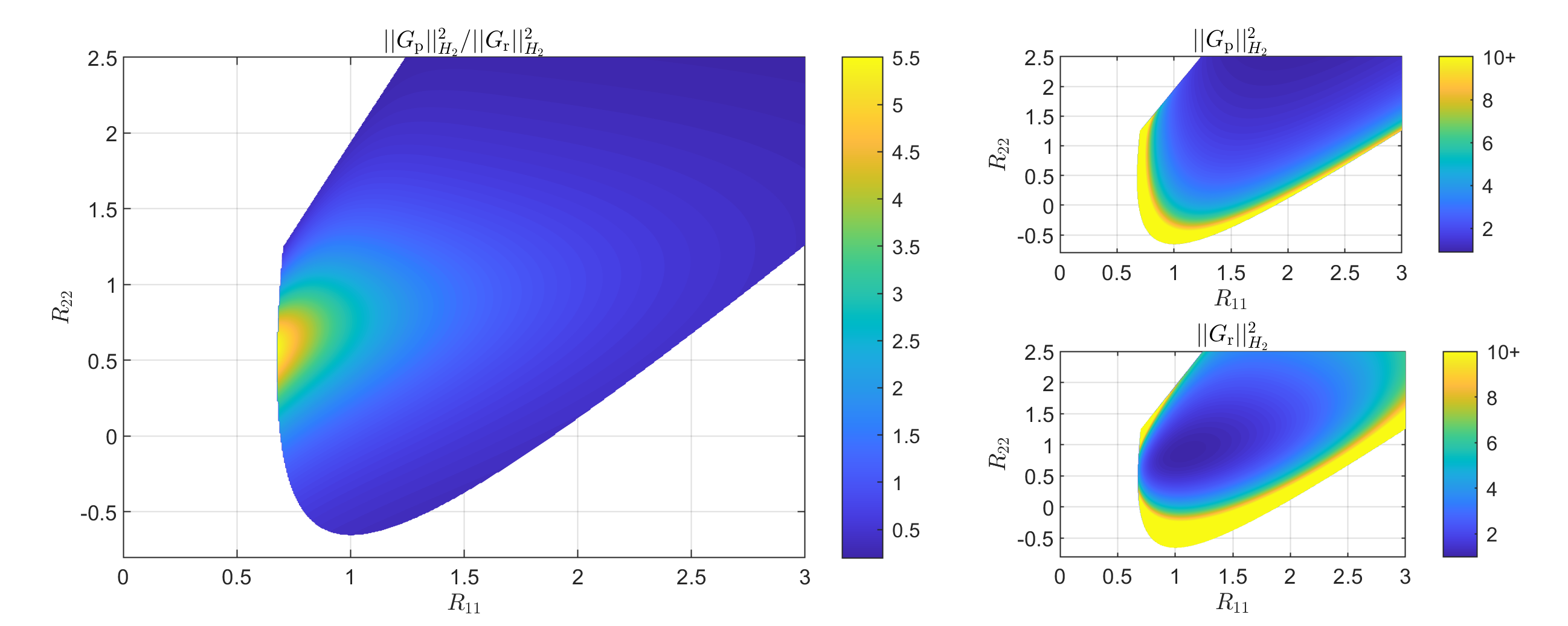}
\caption{The effect of multiplicative attacks with diagonal $R= \begin{bmatrix}
R_{11}&0\\
0&R_{11}\\
\end{bmatrix}\in S^{n_y}$. Left figure: the ratio $\frac{\Vert G_{\rm p}\Vert^2_{H_2}}{\Vert G_{\rm r}\Vert^2_{H_2}}$ is shown as a function of $R_{11}$ and $R_{22}$. The colored area represents the values of $(R_{11}, R_{22})$ for which the matrix $A_R$ is stable. The point $(1,1)$ corresponds to the attack-free case. The maximum impact occurs at $(0.685, 0.56)$, where the ratio reaches 5.35. Right figures: the corresponding values of $\Vert G_{\rm p}\Vert^2_{H_2}$ and $\Vert G_{\rm r}\Vert^2_{H_2}$ are shown separately.}
\label{fig_mult_all}
\end{figure*}
\begin{Theorem}
\label{thm:ratio_bound2}
 Assume that $B$ has full column rank, and that $R\in \mathcal{R}_{\rm s}^{n_y}$ is such that
 $$
 {\rm Re}\,\lambda_{\rm max}(A_R)\le -\alpha_*<0 
 $$
 for some $\alpha_*>0$.
Then the performance ratio \eqref{max_prob} satisfies 
\begin{equation}\label{eq:ratio_bound_Th2}
\frac{\tr\{C_{\rm p}P_R C_{\rm p}^{\rm T}\}}{\tr\{C_{{\rm r},R} P_R C_{{\rm r},R}^{\rm T}\}}
\;\le\;
8 \frac{M_*^2 \left\Vert A_R\right\Vert_2 \, \left\Vert B\right\Vert_2^2}{\sigma_{\min}^2(B) \alpha_*} \, \frac{\|C_{\rm p}\|_F^2}{\|E_R\|_F^2+\|C_{{\rm r}}\|_F^2},
\end{equation}
where $M_*$ is the semigroup bound such that 
\begin{equation}\label{smgrp}
\|e^{A_R t}\| \le M_* e^{-\alpha_* t}.
\end{equation}
\end{Theorem}
\begin{proof}
See Appendix.
\end{proof}

\begin{Remark}
This bound provides an explicit estimate of the worst-case performance ratio \eqref{max_prob} in terms of system parameters, semigroup bounds, and the multiplicative attack $R$, and can be interpreted as an upper bound on the corresponding $H_2$-norm ratio between the performance and residual outputs under stabilizing attacks.
\end{Remark}

\begin{Remark}
From a structural perspective, multiplicative attacks are most effective when the closed-loop system exhibits slow or marginally stable modes that are controllable and observable through the performance output. Denoting such a mode by \(v\), the attack maximizes the performance-to-residual energy ratio when \(v\) lies approximately in the nullspace of the residual map \(C_{{\rm r},R}\) while remaining visible to the performance output. In this scenario, the residual energy remains bounded, whereas the performance output can accumulate substantially, representing a stealthy attack. Non-normality of the closed-loop system can further amplify energy growth along certain directions.

Formally, an attack is maximally effective when there exists a closed-loop mode \(v\) satisfying
\begin{equation}\label{eq:hidden_mode}
C_{{\rm r},R} v \approx 0, \quad C_{\rm p} v \neq 0.
\end{equation}
Explicitly, for the residual map
$C_{{\rm r},R}$,
a hidden mode requires
\begin{equation}\label{eq:residual_cancellation}
(R-I) C_{m,o} v_1 + C_{m,o} v_2 \approx 0,
\end{equation}
with \(v = [v_1^{\rm T}, v_2^{\rm T}]^{\rm T}\) in the extended state coordinates, while the performance output satisfies
\begin{equation}\label{eq:performance_visibility}
C_{\rm p} v = (C_{\rm p,o}+D_{\rm p,o}L) v_1 -D_{\rm p,o}L v_2 \neq 0.
\end{equation}
These conditions define a structural alignment that allows the performance output to grow while the residual remains small.
However, we can note that achieving a perfectly stealthy multiplicative attack, one that simultaneously cancels the residual, excites the performance output, and exploits a slow or marginally stable mode, is extremely challenging in practice. Specifically, such an attack would require designing $R$ so that $A_R$ has a marginally stable eigenmode $v$ that simultaneously satisfies \eqref{eq:residual_cancellation} and \eqref{eq:performance_visibility}. These conditions are tightly coupled and highly nonlinear in $R$, making exact residual cancellation difficult to achieve. This observation further motivates our assumption of a stability margin in Theorem~\ref{thm:ratio_bound2}, which excludes slow or marginally stable modes and ensures that the worst-case performance ratio remains finite and well-posed.
\end{Remark}

%\begin{Prop}\label{prop_H2_2_LMI}
%For each fixed matrix $R\in S^{n_y}$ and $T\to\infty$, the ratio $\frac{\Vert G_{\rm p}\Vert^2_{H_2}}{\Vert G_{\rm r}\Vert^2_{H_2}}$
%can be calculated as $\frac{\tr\left\{C_{\rm p} X_{\rm p}C_{\rm p}^{\T}\right\}}{\tr\left\{C_{{\rm r}, R} X_{\rm r}C_{{\rm r}, R}^{\T}\right\}}$, where
%$X_{\rm p}$ and $X_{\rm r}$  are the solutions of the following optimization problems
%\begin{equation}\label{Opt1_2}
%\begin{aligned}
%& \inf\limits_{X_{\rm p}, \gamma_{\rm p}}\gamma_{\rm p}, \quad \mbox{s.t.}\quad \gamma_{\rm p}>0, \quad X_{\rm p}=X_{\rm p}^{\T}>0, \\
%& \begin{bmatrix}
%X_{\rm p}&X_{\rm p}C_{\rm p}^{\T}\\
%C_{\rm p}X_{\rm p}&\gamma_{\rm p}I\\
%\end{bmatrix}>0, \,\, \begin{bmatrix}
%X_{\rm p}A_R^{\T}+A_RX_{\rm p}&B\\
%B^{\T}&-I\\
%\end{bmatrix}<0,  
%\end{aligned}
%\end{equation}
%and
%\begin{equation}\label{Opt2_2}
%\begin{aligned}
%& \inf\limits_{X_{\rm r}, \gamma_{\rm r}}\gamma_{\rm r}, \quad \mbox{s.t.}\quad \gamma_{\rm r}>0, \quad X_{\rm r}=X_{\rm r}^{\T}>0, \\
%& \begin{bmatrix}
%X_{\rm r}&X_{\rm r}C_{{\rm r}, R}^{\T}\\
%C_{{\rm r}, R}X_{\rm r}&\gamma_{\rm r}I\\
%\end{bmatrix}>0, \,\, \begin{bmatrix}
%X_{\rm r}A_R^{\T}+A_RX_{\rm r}&B\\
%B^{\T}&-I\\
%\end{bmatrix}<0,  
%\end{aligned}
%\end{equation}
%respectively.
%\end{Prop}

%%%%%%%%%%%%%%%%%%%%%%%%%%%%%%%%%%%%%%%%%%%%

\section{Numerical Example}
In this section, we will demonstrate, how multiplicative attacks can affect  the system performance.
Consider the system \eqref{sys_lin_per}, \eqref{obs_lin} with the following parameters
\begin{equation}\label{sys_ex}
\begin{aligned}
A_{\rm p} & = \begin{bmatrix}
1&-2&-1\\
0&-0.5&0\\
0&0&-0.1\\
\end{bmatrix},\,\, 
B_{\rm p}= \begin{bmatrix}
0\\
1\\
1\\
\end{bmatrix},\,\, 
B_{\rm w}= I,\\
C_{\rm m, o} & = \begin{bmatrix}
1&0&0\\
0&0&1\\
\end{bmatrix},\,\, 
C_{\rm p, o}= \begin{bmatrix}
0&1&0
\end{bmatrix},\,\, 
D_{\rm p, o}=0,
\end{aligned}
\end{equation}
i.e., $x=\left[x_1,x_2,x_3\right]^{\T}$, $y_{\rm p} = x_2$, $y_{\rm m} =  \left[x_1,x_3\right]^{\T}$.
The controller and observer gains are chosen as
$$
L=[2.43, -3.24, -0.66],\quad K=\begin{bmatrix}
3&0&0\\-1&0&0.9\\
\end{bmatrix}^{\T},
$$
such that the eigenvalues of the matrices $A_{\rm p}+B_{\rm p}L$ and $A_{\rm p}-KC_{\rm m,o}$ are $\left\{-1,-2,-0.5\right\}$.

\begin{figure}
\begin{center}
\subfloat[{\scriptsize The ratio $\frac{\tr\left\{ C_{\rm p}PC_{\rm p}^{\T}\right\}}{\tr\left\{ C_{{\rm r}, R}PC_{{\rm r}, R}^{\T}\right\}}$ with $P$ as a function of $T$, see \eqref{Gram}.}]{
\includegraphics[width=3.45in]{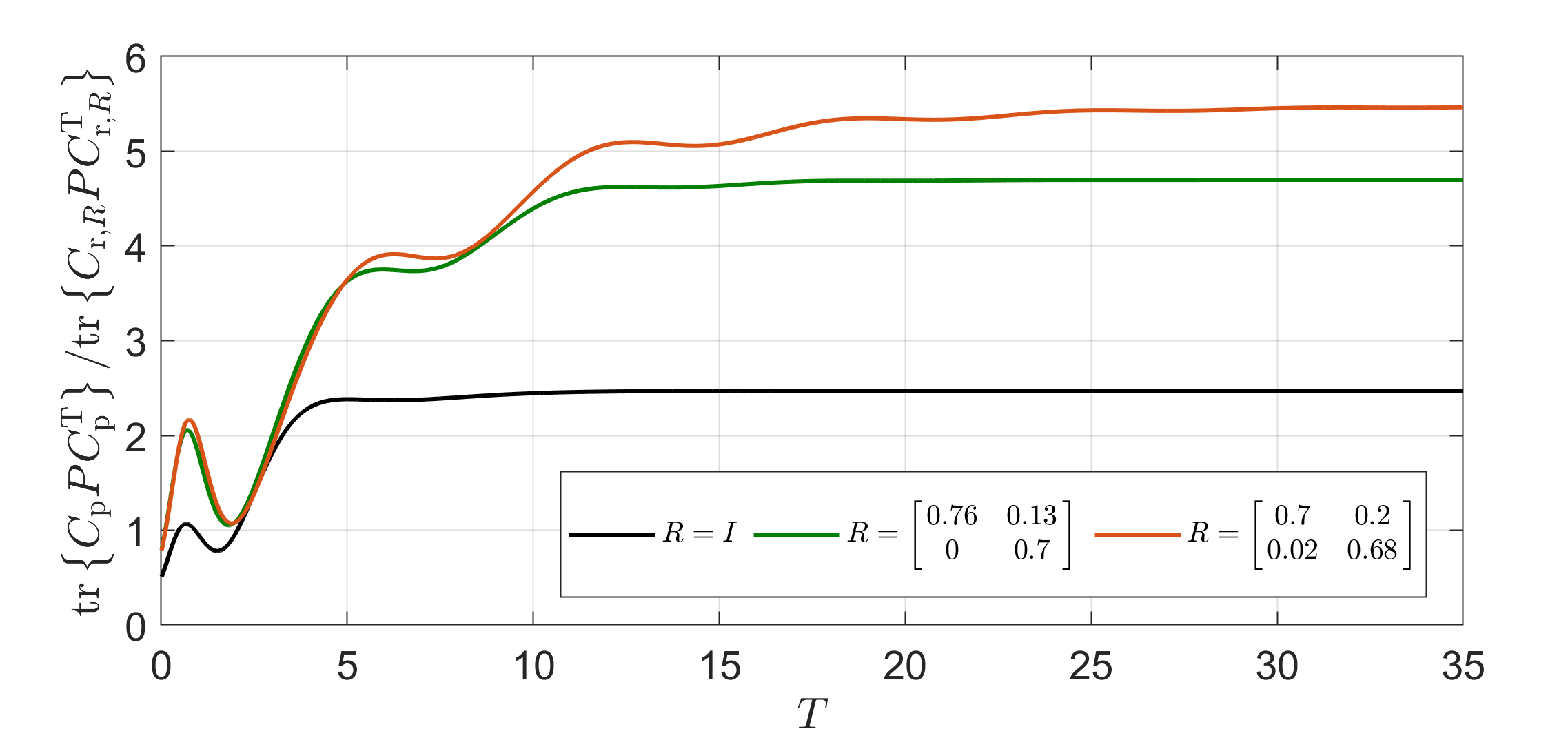}
\label{ratio_traj}
}

\subfloat[{\scriptsize Solid lines illustrate the performance output energy, while dashed lines show the \\ residual output energy.}]{
\includegraphics[width=3.45in]{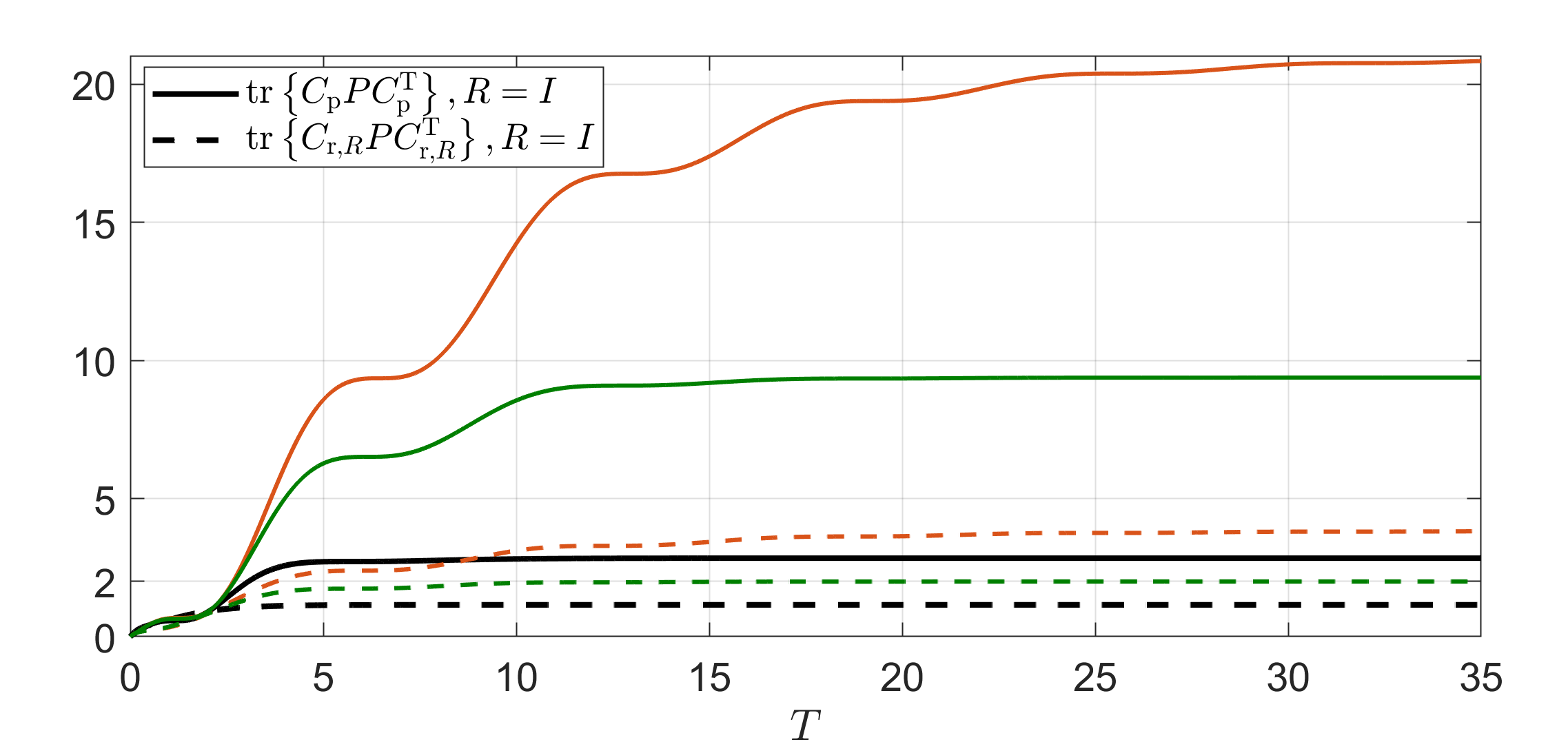}
\label{abs_traj}
}
\end{center}
\caption{The effect of multiplicative attacks for different $R$. The black lines ($R=I$) represent the attack-free scenario. The red lines illustrate the worst case over all $R\in S^{n_y}$, where the ratio $\frac{\Vert G_{\rm p}\Vert^2_{H_2}}{\Vert G_{\rm r}\Vert^2_{H_2}}=\lim\limits_{t\to\infty}\frac{\tr\left\{ C_{\rm p}PC_{\rm p}^{\T}\right\}}{\tr\left\{ C_{{\rm r}, R}PC_{{\rm r}, R}^{\T}\right\}}$  reaches its maximum. The green lines correspond to a stealthy attack, where  $R\in S^{n_y}$ is constrained by $\Vert G_{\rm r}\Vert^2_{H_2}<2$.}
\label{mult_traj}
\end{figure}

We first consider multiplicative attacks with a diagonal matrix
$
R= \begin{bmatrix}
R_{11}&0\\
0&R_{11}\\
\end{bmatrix}.
$
Figure~\ref{fig_mult_all} illustrates the values of the ratio $\frac{\Vert G_{\rm p}\Vert^2_{H_2}}{\Vert G_{\rm r}\Vert^2_{H_2}}$ for different $R_{11}$ and $R_{22}$ such that the matrix $A_R$ is stable. The maximum impact occurs for $R_{11}=0.685$ and $R_{22}=0.56$, with the ratio value being $5.35$. In contrast, the maximum impact for the attack-free case ($R=I$) is $2.46$.
For arbitrary matrix $R\in S^{n_y}$, the adversary can achieve a slightly higher impact than in the diagonal case by selecting
$
R= \begin{bmatrix}
0.7&0.2\\
0.02&0.68\\
\end{bmatrix}
$, which yields a ratio value of $5.46$.
At the same time, the corresponding energy of the residual output, $\Vert G_{\rm r}\Vert^2_{H_2}$, is $3.81$, which may trigger a detection alarm if the threshold is set below this value.
If we consider stealthy attacks only, i.e., by limiting $R$ to the class $\Vert G_{\rm r}\Vert^2_{H_2}<\varepsilon_{\rm tr}$ (and choose, e.g., $\varepsilon_{\rm tr}=2$), we obtain that the maximal ratio $\frac{\Vert G_{\rm p}\Vert^2_{H_2}}{\Vert G_{\rm r}\Vert^2_{H_2}}$ is $4.65$ for $
R= \begin{bmatrix}
0.76&0.13\\
0 &0.7\\
\end{bmatrix}
$, as illustrated in Figure~\ref{mult_traj}.
Therefore, the presence of multiplicative attacks can significantly degrade performance. Notably, since the obtained matrices $R$ exhibit strict diagonal dominance with elements around 0.7, the maximal effect is achieved by simply reducing the power of the measurement signal by approximately 30\%.

%\begin{figure}
%\centering
%\includegraphics[width=3.7in]{./fig/02_ratio.png}
%\caption{}\label{delta_max_all}
%\end{figure}

%%%%%%%%%%%%%%%%%%%%%%%%%%%%%%%%%%%%%%%%%%%%%%%%%%%%%%%%%%%%%%%%%%%%%%%%%%%%%%%%
\section{Conclusions}

This paper presented an $H_2$-norm approach to the performance analysis of NCSs under multiplicative routing transformations. By modeling routing manipulations as structured, energy-preserving modifications of the measurement pathway, we characterized their effect on closed-loop behavior through the ratio between the performance and residual output energies. Equivalent LMI-based formulations and analytical upper bounds were derived, enabling efficient evaluation of worst-case performance degradation and providing structural insight into stealthy routing attacks.

This $H_2$-based formulation provides a tractable and physically meaningful characterization of how the routing matrix $R$ alone affects performance and detectability. The extension of this framework to additive or time-varying attacks, which require external energy and are more suitably analyzed via $H_\infty$-methods, will be the subject of future work.

Future work may also consider mitigation strategies, such as structural design of controller and observer gains or the use of small private excitation signals to prevent multiplicative-only manipulations from canceling residual contributions and to enhance robustness against stealthy routing transformations.

\bibliographystyle{IEEEtran}
\bibliography{references}

\appendix

{\bf Proof of Proposition~\ref{prop_H2}.}
We first establish the equivalence between problems~\eqref{max_prob01} and~\eqref{max_prob2}.
Since $A_R$ is stable, the process $y_{\rm p}(t)$ converges to a stationary one, and
\[
\E\!\left\{\lim_{t\to\infty} y_{\rm p}^{\T}(t)y_{\rm p}(t)\right\}
= \frac{1}{2\pi} \int_{-\infty}^{\infty} 
\tr\!\left\{ S_{y_{\rm p}}(\omega) \right\} d\omega,
\]
where $S_{y_{\rm p}}(\omega)$ denotes the spectral density of $y_{\rm p}(t)$.  
Since $y_{\rm p} = G_{\rm p}w$ with $S_w(\omega) = I$, the spectral density can be expressed as
\[
S_{y_{\rm p}}(\omega) = G_{\rm p}(i\omega)S_w(\omega)G_{\rm p}^*(i\omega)
= G_{\rm p}(i\omega)G_{\rm p}^*(i\omega).
\]
An analogous expression holds for $y_{\rm r}(t)$, which immediately implies the equivalence in between \eqref{max_prob01} and \eqref{max_prob2}.

Next, we show that~\eqref{max_prob2} and~\eqref{max_prob} are equivalent for $T\to\infty$.  
The impulse response from $w$ to $y_{\rm p}$ is given by
\begin{equation*}\label{imp}
h_{\rm p}(t) = \int_0^t C_{\rm p}\e^{A_R^{\T}(t-\tau)}B\,\delta(\tau)d\tau 
= C_{\rm p}\e^{A_R^{\T}t}B.
\end{equation*}
Its $L_2$-norm is
\begin{equation*}\label{imp2}
\begin{aligned}
\|h_{\rm p}\|_{L_2}^2 
&= \int_0^{\infty} \tr\!\left\{ h_{\rm p}^{\T}(t)h_{\rm p}(t) \right\} dt
= \int_0^{\infty} \tr\!\left\{ h_{\rm p}(t)h_{\rm p}^{\T}(t) \right\} dt \\
&= \int_0^{\infty} 
\tr\!\left\{ C_{\rm p}\e^{A_R t}BB^{\T}\e^{A_R^{\T}t}C_{\rm p}^{\T} \right\} dt.
\end{aligned}
\end{equation*}
On the other hand, by Parseval’s theorem,
\begin{equation*}\label{imp3}
\|h_{\rm p}\|_{L_2}^2 
= \frac{1}{2\pi}\int_{-\infty}^{\infty} 
\tr\!\left\{ G_{\rm p}(i\omega)G_{\rm p}^*(i\omega) \right\} d\omega
= \|G_{\rm p}\|_{H_2}^2.
\end{equation*}
Therefore,
\begin{equation*}\label{opt1_pf}
\|G_{\rm p}\|_{H_2}^2 
= \tr\!\left\{ C_{\rm p}P_R C_{\rm p}^{\T} \right\},
\end{equation*}
where $P_R$ is defined by \eqref{Gram} for $T\to\infty$.  
The same reasoning applies to $y_{\rm r}(t)$, which establishes the equivalence between~\eqref{max_prob2} and~\eqref{max_prob}.  
Hence, Proposition~\ref{prop_H2} follows.

%%%%%%%%%%%%%%%%%%%%%%%%%%%
{\bf Proof of Proposition~\ref{prop_H2_LMI_Boyd_contr}.}
Since $A_R$ is stable and $T\to\infty$, the matrix $P_R$ can be equivalently obtained as the solution of the Lyapunov equation
\begin{equation}\label{lyap_eq_Boyd}
P_RA_R^{\T}+A_R P_R=-BB^{\T}.
\end{equation}
Consider also the associated Lyapunov inequality
\begin{equation}\label{lyap_ineq_Boyd}
 XA_R^{\T}+A_R X\le-BB^{\T}. 
\end{equation}
From \eqref{lyap_eq_Boyd} and \eqref{lyap_ineq_Boyd}, it follows that
\begin{equation}\label{lyap_ineq2_Boyd}
( X-P_R)A_R^{\T}+A_R( X-P_R)\le0, 
\end{equation}
and, since $A_R$ is stable, we have $ X - P_R \ge 0$.  
This implies
$
C_{\rm p} XC_{\rm p}^{\T}\ge C_{\rm p} P_RC_{\rm p}^{\T}.
$
and hence
$
\tr\left\{C_{\rm p} XC_{\rm p}^{\T}\right\} \ge \tr\left\{C_{\rm p} P_RC_{\rm p}^{\T}\right\}.
$
Therefore, the following optimization problem holds:
\begin{equation}\label{opt1a_pf_Boyd}
\tr\left\{C_{\rm p} XC_{\rm p}^{\T}\right\} \to \inf, \quad  XA_R^{\T}+A_R X<-BB^{\T},  \,\,X>0.
\end{equation}
This can be equivalently written as
\begin{equation}\label{opt1b_pf_Boyd}
\begin{aligned}
&\gamma_{\rm p}  \to \inf, \quad \tr\left\{Z_{\rm_p}\right\}<\gamma_{\rm_p}, \quad X>0, \quad \gamma_{\rm p}>0,\\
&C_{\rm p} XC_{\rm p}^{\T}<Z_{\rm p}, \quad XA_R^{\T}+A_R X<-BB^{\T}.
\end{aligned}
\end{equation}
Finally, applying the Schur complement yields the LMI representation~\eqref{Opt1_Boyd}, which completes the proof.

%%%%%%%%%%%%%%%%%%%%%%%%%%%%%%

{\bf Proof of Proposition~\ref{prop_H2_2}.}
For $R\in \mathcal{R}_{\rm s}^{n_y}$ and $T\to\infty$, the matrix $P_R$ is a unique solution of the Lyapunov equation
$$
 XA_R^{\T}+A_R X=-BB^{\T},
$$ and
the optimization problem \eqref{max_prob} can be equivalently rewritten as
$$
\sup\limits_{R\in \mathcal{R}_{\rm s}^{n_y}, \alpha\ge0}\alpha,\quad\mbox{s.t.} \,\,\,\tr\left\{C_{\rm p}XC_{\rm p}^{\T}\right\}\ge\alpha \tr\left\{C_{{\rm r}, R}XC_{{\rm r}, R}^{\T}\right\}.
$$
Using the cyclic property of the trace, we finally obtain
$$
\begin{aligned}
&\tr\left\{C_{\rm p}XC_{\rm p}^{\T}\right\}-\alpha \tr\left\{C_{{\rm r}, R}XC_{{\rm r}, R}^{\T}\right\}
=
\tr\left\{XC_{\rm p}^{\T}C_{\rm p}\right\}\\
&-\alpha \tr\left\{XC_{{\rm r}, R}^{\T}C_{{\rm r}, R}\right\}
=\tr\left\{X\left(C_{\rm p}^{\T}C_{\rm p}-\alpha C_{{\rm r}, R}^{\T}C_{{\rm r}, R}\right)\right\}.
\end{aligned}
$$

%%%%%%%%%%%%%%%%%%%%%%%%%%%%%%%%
{\bf Proof of Theorem~\ref{thm:ratio_bound2}.}
Along the lines of the proof of Theorem~\ref{thm:ratio_bound}, we arrive at the inequality \eqref{eq:rat_bounds}. To estimate the condition number $\kappa(P_R)$, using \eqref{smgrp}, we first obtain
\[
\lambda_{\max}(P_R) \le \int_0^\infty \left\Vert \e^{A_R t} B B^{\rm T} \e^{A_R^{\rm T} t}\right\Vert_2 dt \le \frac{M_*^2}{2 \alpha_*} \left\Vert B\right\Vert_2^2.
\]
Then we need to obtain the lower bound for $\lambda_{\rm min}(A_R)$.
Take any unit vector $x  \in \mathbb{R}^{2n_x}$. Then
$
\lambda_{\min}(P_R) 
= \min_{\|x\|_2 = 1} x^T P_R x.
$
We have that
\[
x^{\rm T} P_R x = \int_0^\infty  \left\Vert B^{\rm T} \e^{A_R^{\rm T} t} x\right\Vert_2^2 dt \ge \int_0^{t_0} \left\Vert B^{\rm T} \e^{A_R^{\rm T} t} x\right\Vert_2^2 dt
\]
for some $t_0>0$. Since $\left\Vert \left(A_R^{\rm T}\right)^k\right\Vert_2\le\left\Vert A_R^{\rm T}\right\Vert_2^k$, by direct calculations we obtain
$
\left\Vert \e^{A_R^{\rm T} t} - I \right\Vert_2 \le \e^{\left\Vert A_R\right\Vert_2t}-1.
$
If we take $t_0=\frac{\ln{3/2}}{\left\Vert A_R\right\Vert_2}$, then $\e^{\left\Vert A_R\right\Vert_2t}-1\le \frac{1}{2}$ for all $t\in[0,t_0]$. 
Thus,
$$
\left\Vert \e^{A_R^{\rm T} t} x \right\Vert_2 \ge \left(1-\left\Vert \e^{A_R^{\rm T} t} - I \right\Vert_2\right)\Vert x\Vert_2\ge\frac{1}{2}\Vert x\Vert_2=\frac{1}{2}.
$$
Since $B$ has full column rank with minimum singular value $\sigma_{\min}(B) > 0$,  
\[
\left\Vert B^{\rm T} \e^{A_R^{\rm T} t} x \right\Vert_2 \ge \sigma_{\min}(B) \left\Vert \e^{A_R^{\rm T} t} x \right\Vert_2 \ge \frac{\sigma_{\min}(B)}{2}.
\]
Integrating over $[0,t_0]$:
\[
x^{\rm T} P_R x \ge t_0 \frac{\sigma_{\min}^2(B)}{4} = \frac{\sigma_{\min}^2(B)}{16 \left\Vert A_R\right\Vert_2}.
\]
Hence,
\[
\lambda_{\min}(P_R) \ge \frac{\sigma_{\min}^2(B)}{16 \left\Vert A_R\right\Vert_2}.
\]
Therefore,
\[
\kappa(P_R) = \frac{\lambda_{\max}(P_R)}{\lambda_{\min}(P_R)} \le 8 \frac{M_*^2 \left\Vert A_R\right\Vert_2 \, \left\Vert B\right\Vert_2^2}{\sigma_{\min}^2(B) \alpha_*}.
\]
leading to \eqref{eq:ratio_bound_Th2}.

\end{document}